\documentclass[letterpaper,twocolumn,10pt]{article}
\usepackage{usenix2023_SOUPS}
\usepackage{amsmath,amssymb,amsfonts}
\usepackage{algorithmic}
\usepackage{graphicx}
\usepackage{textcomp}
\usepackage{xcolor}
\usepackage{xspace}
\usepackage{booktabs}
\usepackage{todonotes}
\usepackage{comment}

\usepackage{tabularx}
\usepackage{adjustbox}
\newcolumntype{R}[2]{%
    >{\adjustbox{angle=#1,lap=\width-(#2)}\bgroup}
    l%
    <{\egroup}%
}
\newcommand*\rot{\multicolumn{1}{R{45}{.8ex}}}
\newcommand{\T}{\(\bullet{} \)}
\newcommand{\C}{\(\circ\)} 

\usepackage{fontawesome}
\newcommand{\exthref}[1]{\xspace\href{#1}{\faExternalLink}}
\newcommand{\shorthref}[1]{\href{#1}{\faExternalLink}}
\newcommand{\fthref}[1]{\footnote{\url{#1}}}

\begin{document}

\date{}

\title{\Large \bf Communicating on Security within Software Development Issue Tracking}

\def\plainauthor{Léon McGregor, Manuel Maarek, Hans-Wolfgang Loidl}

\author{%
{\rm Léon McGregor}\\
Heriot-Watt University\\
lm356@hw.ac.uk\\
\and
{\rm Manuel Maarek}\\
Heriot-Watt University\\
M.Maarek@hw.ac.uk\\
\and
{\rm Hans-Wolfgang Loidl}\\
Heriot-Watt University\\
H.W.Loidl@hw.ac.uk\\
} 

\maketitle
\thecopyright

\begin{abstract}
    During software development, balancing security and non security issues is challenging.
    We focus on security awareness and approaches taken by non-security experts using software development issue trackers when considering security.
    We first analyse interfaces from prominent issue trackers to see how they support security communication and how they integrate security scoring.
    Then, we investigate through a small scale user study what criteria developers take when prioritising issues,
    in particular observing their attitudes to security.

    We find projects make reference to CVSS summaries (Common Vulnerability Scoring System), often alongside CVE reports (Common Vulnerabilities and Exposures),
    but issue trackers do not often have interfaces designed for this.
    Users in our study were not comfortable with CVSS analysis,
    though were able to reason in a manner compatible with CVSS.
    Detailed explanations and advice were seen as helpful in making security decisions.
    This suggests that adding improvements to communication through
    CVSS-like questioning in issue tracking software can elicit better security interactions.
\end{abstract}

\section{Introduction}

Discrepancies exist between what security experts desire and the guidance developers end up following~\cite{Maz_P2ASCCCS-2022},
so building better security practices into software development is key for exposing security to non-experts.
Modern Development Operations (DevOps) processes,
streamline tracking and implementing features, bug fixes, and
management of issues arising in software development~\cite{WFW+_Q-2019}.
A key concern for security is whether,
amongst the many DevOps security practices~\cite{UW_PIWCSED-2016},
processes include prioritising reported security issues.
Prioritisation of issues during development is an important step,
as it dictates the approach that a whole project will take towards developing their product.
Priority tensions in a project mean that many different aspects will compete for priority,
and have an impact on developer's approaches to security and analytical thinking.
``Consideration of revenue is rational'' for developers to prioritise~\cite{CHP+_2ISDCS-2021},
so care is needed to balance security against functional requirements.
In order for an application to be secure,
security issues ought to get a high priority.
Making this decision well requires project members to be informed on security risks and to be involved in the decision making process, beyond just the security experts.

Investigating whether tools used during DevOps 
do enable security is important
as there are many factors that impact adoption~\cite{XWM_P1ACCSCWSC-C1-2014}.
DevOps must also enable security professionals to use
``communication and methods by which [non-security developers] already share knowledge as part of their workflow''~\cite{AO_NSPW2-2020}. 
When considering security,
communication and management during development are key aspects to improving security motivation~\cite{BBH+_IaST-2008}.
There is a need to investigate if DevOps processes and tools encourage good security among non-security experts.

Our investigation is twofold.
First, we survey the design and security approaches within large projects which track issues openly (Section~\ref{pri:tools}).
Then, we conduct a developer study observing non-security experts approaches to security prioritisation of issues and perception of the role of CVSS (Common Vulnerability Scoring System)~\cite{FIR_F--FoIRaST-2019} as a security analysis system (Section~\ref{devstudy}).
We frame our work around the following research questions.

\noindent
\textbf{RQ1} How do software development issue tracking systems integrate security considerations?
\\\noindent
\textbf{RQ2} Does prompting for CVSS during issue management have the potential to be a useful interaction?
\\\noindent
\textbf{RQ3} How can non-security experts better engage with security during project issue management?

\paragraph{Contributions}
Our survey covers four large open software development issue trackers and their usage in openly tracked projects (Section~\ref{pri:tools} details our selection criteria).
Then we run a developer study investigating how non-security-experts make security decisions.
Four participants engaged, making it a scale small investigation,
but we nevertheless draw the following preliminary findings which form the main contributions of this paper.
1) Existing software development issue tracking tools lack the design to fully convey security concerns.
2) Non-experts seem not comfortable with using CVSS analysis.
3) However, CVSS seems to be considered helpful by a non-security expert/experienced project manager to prioritise security-impacting issues.
4) Security inquiry through questioning and sharing advice could make security more accessible to non-experts.
5) Security related metadata could be integrated into issue trackers, elicited by text answers to security questions, or optionally CVSS scores.

\vspace{-0.2cm}
\paragraph{Security classifications and scoring}

Numerous methods have been adopted across the security industry to help classify the impact of security flaws.
CWE identifiers (Common Weakness Enumeration)~\cite{MIT_-2006} represent common classes of bug with similar behaviours.
They can be assigned to a bug to better describe what the problem is in relation to others.
CVE identifiers (Common Vulnerabilities and Exposures)~\cite{MIT_-1999} are assigned to specific cases of flaws
and uniquely identify a single case of a security vulnerability or exposure of a sensitive system.
CVSS scores are values given to a reported vulnerability,
calculated by measuring the impact of a security flaw across several dimensions.
CVE records contain CVSS analysis, and many CVE records reference CWE classifications.
CVSS analysis can either be shared as a score between 0 and 10 (most severe) or
as a string which encodes all of the individual components of the scoring metrics,
allowing individuals to see precisely what the risks are.
CVSS is a method of measuring the severity of a security issue, and is often attached to a specific CVE,
though there is nothing that precludes it being used outside of CVE only.
Some research suggests there is significant disagreement amongst security experts over the individual scores that might be generated through CVSS~\cite{HA_C&S-2015},
however there are findings suggesting that in general the CVSS scores produced by databases are trustworthy~\cite{JLEF_ITDSC-2018}.
This tension suggests that these scoring mechanisms are worth investigating further, particularly with respect to non-security experts.
There exist other classification and scoring systems,
such as CWSS scoring system~\cite{MIT_-2014}, similar to CVSS, but not focused on specific security incidents.
We focus on CVSS as it was specifically designed to emphasise and rank which bugs should be prioritised for patching~\cite{pendletonSurveySystemsSecurity2016}.

\section{Issue Tracker Survey}\label{pri:tools}

\newcommand{\bugger}[2]{Bug~\texttt{\href{#1#2}{#2}}}
\newcommand{\bugz}[1]{\bugger{https://bugzilla.mozilla.org/show_bug.cgi?id=}{#1}}
\newcommand{\bugm}[1]{\bugger{https://bugs.chromium.org/p/chromium/issues/detail?id=}{#1}}
\newcommand{\bugj}[1]{\bugger{https://jira.atlassian.com/browse/}{#1}}
\newcommand{\bugg}[1]{\bugger{https://gitlab.com/gitlab-org/gitlab/-/issues/}{#1}}

To answer our research questions we investigate the design characteristics and security approaches of platforms,
before we conduct a developer study investigating developer approaches within issue trackers.
In this section, we investigate issue tracking tools in order to see how well they integrate and promote security.
We have decided to analyse the public facing instances of issue trackers for projects run by the same groups that created the bug trackers,
assuming that these developers would be the most likely to fully utilise the capabilities of their issue tracking tools.
Note that we are not focused on performing in-depth ethnographic analysis of these projects,
but simply interested to see how they publicly use their own issue tracking platforms in relation to security.
Platforms referenced here were accessed in July 2022.

\vspace{-0.2cm}
\paragraph{Investigation steps}

Our investigation followed steps for each issue tracker:
1) Explore and identify all the interface elements,
2) Find security issues reported within the project and observe interesting interactions,
3) Find mentions of CVE, CVSS, CWE, or discussion relating to security choices.

Interface exploration involved identifying common and differing interface elements on the individual bug/issue pages
and seeing whether the projects made use of these available elements.
Security issues needed to be located by a broad search for ``security'' in body text or labelling
and identifying how  particular project demarcates security issues,
then doing a deeper investigation on terms such as ``CVE'' or ``CVSS''.

\vspace{-0.2cm}
\paragraph{Selection criteria}

There are many DevOps, issue tracking, and project management tools,
so we narrowed down to a specific subset with publicly visible trackers to allow for the focused analysis detailed above.
We selected projects and tracking systems
when the system would
1) Track issues in a software development project,
2) Not require a special plugin or extension,
3) Have a public tracker used by the manufacturers,
4) Serve large projects.

We settled on the following 4 issue trackers:
Bugzilla (Firefox Web Browser), 
Monorail (Chrome Web Browser), 
Jira (Atlassian DevOps Tools), 
GitLab. 
Among other major open systems, we discarded GitHub as it does have one central bug tracking system, and
Linux Kernel development
as it relies on mailing lists
which are out of scope of this review.
We note other platforms (Trac, Redmine) met the criteria but these were not selected for full investigation.

\subsection{Investigation}

\paragraph{Bugzilla}
Bugzilla is a bug tracker developed by Mozilla for use with developing the Firefox browser.
Within the Firefox project, the project uses all of the fields that are available in the interface.
In addition, Bugzilla allows site administrators to add custom fields and properties
if the project managers feel they would benefit from it%
\exthref{https://bugzilla.readthedocs.io/en/5.0.4/administering/custom-fields.html}.

Bugzilla does not include any built-in support for tracking or measuring CVSS scores,
but there are cases where bug report participants have included comments that reference CVSS scores.
From searching the Firefox Bugzilla instance for references to CVSS,
we find that most of the bugs that include a CVSS score are CVEs which have been copied to the Mozilla-run bug tracker from other sources.
\bugz{1246014}
is a good example of this.
Coming via email from an external security investigator,
the CVSS and related analysis needs to be included in a comment to the bug, as there is no standard field in the UI to place it.

We see that when a CVE is reported, it will include a CVSS score, describing the severity of the bug.
In this case the bug was given a keyword, ``sec-critical'', as a way of tracking the severity.
What is interesting in this case is that the dedicated fields for bug priority and severity were not changed.
This behaviour is seen in other CVEs reported to the Mozilla tracker in this fashion, including
\bugz{814001} and 
\bugz{805121}. 
Some CVEs reported as bugs do however get the relevant fields set in addition to a priority keyword, including
\bugz{1274637} and 
\bugz{1631618}. 
Outside of CVE and CVSS related bugs,
keywords, severity and priorities are used to keep track of bugs,
as can be seen in \bugz{1538007}.
It is unclear just from these observations if there is a consistent or formal procedure that Mozilla has for assigning security priorities in their bug tracker,
but it does indicate that a standardised interface that can keep track of CVSS scores in addition to or in place of many separate labels might be useful.

By performing a broad search across the entire bug tracker for mentions of CVSS
excepting when used to describe CVEs,
we find that no bugs use CVSS scores despite the intended use of CVSS as a means of analysing security bugs.

\paragraph{Monorail}
Monorail is a bug tracker developed by Google,
for use in developing Chrome,
the world's largest web browser.
One differentiating aspect to Bugzilla
issue tracking is the use of labels to track different aspects,
rather than using individual fields%
\exthref{https://chromium.googlesource.com/infra/infra/+/main/appengine/monorail/doc/userguide/concepts.md\#Issue-fields-and-labels}.
Labels can be related using dashes \texttt{-} as an alternative to defining specific fields,
such as \texttt{Target-102},
which indicates that the team aims to resolve an issue by the time version \texttt{102} is released.
This scoped labelling approach might offer more autonomy for individual project members to categorise and prioritise issues than having to rely on a project administrator to add specific fields.
This capability is used frequently within the Chrome project.

Searching for CVSS scores\exthref{https://bugs.chromium.org/p/chromium/issues/list?q=cvss&can=1}
given in bugs reported in the Chromium project tracker reveals a similar pattern to that found in the Mozilla tracker.
Comments and descriptions of bugs only tend to mention CVSS scores when they were originally formed as CVEs, cross-posted to the public tracker.
An illustrative example is found in \bugm{1313172}.
The interface includes a priority field, but in contrast no severity field,
and instead severity labels are used.
Monorail does not have a built-in CVSS field or method to track this
so when required the comments section has to be used to display CVSS scores.

The Chrome project has over 1000 bugs referencing CVSS,
but only 2 confirmed bugs that reference CVSS scores without also mentioning CVE%
\exthref{https://bugs.chromium.org/p/chromium/issues/list?q=cvss\%20-cve&can=1}.
These are
\bugm{571480} and 
\bugm{695474}. 
Both have a priority, but only the first has a security severity label.
This may indicate that there may either be limited utility or a lack of recognition of these metrics outside of the context of CVE reporting.

\paragraph{Jira}
Jira is a bug tracker created by Atlassian.
It is used by many projects including MongoDB and Qt.
Atlassian use Jira as their public facing issue tracker for their own products, which we analyse here.
In a Jira issue many fields, properties, and links are present similar to other issue trackers.
Jira also allows Jira administrators to change the fields available%
\exthref{https://support.atlassian.com/jira-cloud-administration/docs/create-a-custom-field/}.
In the open source project Qt, we see that their issue tracker only has the default fields,
and so is missing the ``severity'' fields present in Atlassian's Jira instance
(seen in \bugger{https://bugreports.qt.io/browse/}{QTBUG-105931}).
This approach means that different projects may approach security in different ways,
depending on how they have configured their setup.

Within Atlassian's public issue tracker there are references to CVSS scores within the comments of bugs%
\exthref{https://jira.atlassian.com/issues/?jql=issuetype\%20\%3D\%20Bug\%20AND\%20text\%20~\%20\%22cvss\%22\%20AND\%20summary\%20!~\%20\%22CVE\%22\%20ORDER\%20BY\%20updated}.
It is interesting that many of the CVSS scores are all presented in the same way,
through a table in comments,
yet Jira's custom fields are not used to standardise this.
We see this presentation added by a `bot' in \bugj{JRASERVER-71198},
indicating desire for some kind of automation for collecting this data.
All the comments in this fashion include a link to an Atlassian CVSS Calculator%
\exthref{https://asecurityteam.bitbucket.io/cvss_v3/}
to explain the severity rating.
The CVSS scores seem to come from CVE notifications which state it
\emph{``is an independent assessment and you should evaluate its applicability to your own IT environment.''}
Whether this phrase is included as a way to instruct the community reading these bugs to be alert,
or simply as a form of legal disclaimer is unclear.
CVSS scores are not limited to just CVE reports,
but also appear in 122 issues unrelated to CVE,
indicating that CVSS identifications bear some importance to the Jira project.
Tracking security is an important part of Atlassian bugs,
across their projects 6646 bugs are related in some way to security,
even excluding CVE%
\exthref{https://jira.atlassian.com/issues/?jql=issuetype\%20\%3D\%20Bug\%20AND\%20resolution\%20in\%20(Unresolved\%2C\%20Fixed\%2C\%20Deployed\%2C\%20Done)\%20AND\%20text\%20~\%20security\%20AND\%20summary\%20!~\%20CVE\%20ORDER\%20BY\%20votes\%20DESC}.

While researching which open source projects use Jira,
we found MongoDB recommends voting as a way to gauge community priorities for fixing some issue.
This is an insight that the addition of a ``voting'' property may be incentivising projects to use it as a way to decide priorities collectively.

\paragraph{GitLab}
GitLab targets the whole DevOps process, including issue tracking.
GitLab is open source and the development team use their own project to track issues.

GitLab's tracker shows many bugs mentioning CVSS without an associated CVE report.
There is a GitLab \bugg{218601}
which is a proposal to standardise the way GitLab tracks CVSS scores,
notably suggesting \emph{``scoring can then be exposed to the user in relevant parts of the UI''.}
GitLab's handbook describes that priorities and severity can be assigned based on the CVSS scores that are generated%
\exthref{https://about.gitlab.com/handbook/engineering/security/threat-management/vulnerability-management/}.
The handbook also suggests that issues with high CVSS scores ought to be labelled as high priority and be mitigated within 24 hours.

For individual projects, GitLab's issue tracking interface does have a weight field.
However, the concrete meaning of this field is not defined and could change project to project.
From GitLab's documentation%
\exthref{https://docs.gitlab.com/ee/user/project/issues/issue_weight.html},
it could refer to any of
\emph{``how much time, value, or complexity a given issue has or costs.''}
This is still an important metric when it comes to prioritisation,
however it is not clear if it alone can be leveraged or relied on to handle
the priorities of security issues.

GitLab already has a CVSS calculator%
\exthref{https://gitlab-com.gitlab.io/gl-security/appsec/cvss-calculator/},
but it is not used for generating scores to prioritise issues.
Instead it is for defining the ``bounty'' to be rewarded for certain bugs.
When issues are reported through the bounty program,
the CVSS scores generated as part of the reporting are converted to priority and severity indicated through labelling.
In some cases the CVSS scores, though used as part of reporting,
are not linked on the issue page itself.
Instead developers need to navigate to a separate page to view it.
An example of this can be see in \bugg{336535}:
\emph{Severity is set as per CVSS calculated on hackerone report.}
Not presenting this inline could add to developer workload.
Designing an interface that can present this information could improve issue prioritisation
to draw developer attention to more severe issues.
Other bugs, such as \bugg{360986}, include basic information of the final CVSS score, and link to more detailed reports again held elsewhere.

GitLab's development team currently tracks CWEs, and other weakness and classifications through labels.
GitLab's team use scoped labels, which allows all CWEs of a specific type to be grouped together%
\exthref{https://gitlab.com/gitlab-org/gitlab/-/labels?subscribed=&search=weakness}.
An open GitLab \bugg{300978}
is an issue discussing a proposal of whether to adopt CWEs as a means of tracking and codifying security issues in GitLab's Secure group.
The evidence of interest in CWE and CVSS indicates intent from GitLab to include more support for these security vulnerability classification systems.

\subsection{Comparison and summary}

\begin{table}
    \caption{Comparison of metadata fields across platforms}
    \begin{tabularx}{\hsize}{Xllll}
        Field & \rot{Bugzilla} & \rot{Monorail} & \rot{Jira} & \rot{GitLab} \\ \toprule
        Priority                        & \T & \T & \T & \C \\
        Severity                        & \T & \C & \T & \C \\
        Weight                          &    &    &    & \T \\
        Votes                           &    &    & \T & \T \\
        Scoped Labels                   &    & \T &    & \T \\
        Milestones                      & \T & \C & \T & \T \\
        Estimated Completion Times      &    & \T &    & \T \\
        Epics                           &    & \C & \T & \T \\
        Colours                         & \T &    & \T & \T \\
        Custom Fields                   & \T & \C & \T &    \\
        CWE                             &    &    &    & \C \\
    \end{tabularx}\label{tab:devops}

Native to all:
Title and Description,
Timestamps,
Component Hierarchy,
Issue types,
Labels,
Attachments,
Related Issues,
Confirmation and Resolution,
Project Member Links
\fbox{%
      Metadata support:
            native fields (\T), possible with labels (\C)
    }
\end{table}

Table~\ref{tab:devops} compares the available features focusing on non-security specific features as none of the investigated issue trackers include dedicated space to discuss security specific concerns.
We see that many features are shared amongst the issue trackers.
There are some outliers.
GitLab seems to offer as many capabilities as is possible.
Where certain features are not present,
many projects using these systems utilise labels or keywords as a means of tracking certain properties of issues.
Only GitLab has a `weight' field,
however its use might be duplicated by severity or estimated completion times in other tracking systems.
Most of the tracking systems make use of colour to draw attention to certain features,
with Bugzilla highlighting important issues in red when in list view,
to GitLab which uses very colourful labels, chosen by project organisers.
Some bug tracking systems have built-in priority tracking fields for their bugs,
and every bug tracking system has some form of keyword or labelling fields.
Developers often seem to use labelling as a means of tracking severity,
suggesting labels may be a more usable interface element.
The use of labelling varies between development projects.

The contrast between Bugzilla and Monorail is interesting when considering severity labelling.
Both Bugzilla and Jira include severity fields, others use labels.
But Firefox development includes a specific keyword despite Bugzilla having a severity field,
and Chrome development uses a severity label as an alternative to a severity field.
This indicates different approaches in how strictly severity is assigned,
and that one single approach is perhaps not suitable for all projects.

The GitLab project tracks certain weaknesses classified by CWE through scoped labelling.
CVSS scores are mentioned within bug tracking for open source projects,
often in relation to externally reported issues such as through CVEs or bug bounties.
The CVSS scores appear within the text of the discussion of the bug,
instead of alongside the labels or fields that describe severity and priority.
This could indicate that though there is a desire for CVSS when discussing critical security bugs like CVEs,
the bug tracker interfaces do not leverage CVSS at all for other kinds of reported security issues,
and offer little support for CVSS when they are used.
Note that CVSS targets reported vulnerabilities
and might not suit all security planning discussions.

In summary, the selected DevOps issue trackers have many commonalities, but also a few differences.
We see wide variation in the approaches taken to labelling, handling severity, and handling references to security issues and reports.
External metrics and classifications are used, but interface support for referencing these is missing,
and interfaces offer no automated support for making decisions based on such metrics.

\section{Developer Study}\label{devstudy}
After investigating how trackers support security,
we focus on how users themselves make security decisions
through our developer study.
The purpose of this study is to
investigate approaches that non-security-experts take when analysing and prioritising security issues.
We examine whether techniques like CVSS analysis are useful,
and what parts of an issue presentation impact most.
This study is conducted online with non-security-expert developers and project managers.

\subsection{Protocol}
The participants will be placed in a fictional setting, a software development company developing financial applications.
When presented with issues,
the participant are asked to consider each one, then assign a priority to it, relative to the others, and answer questions about CVSS scoring.
All participants see the same issues, but in a random order.
The prioritisation will be a scale where the top issue needs to be dealt with first,
and both security and business-critical issues are rated together,
to simulate the prioritisation stresses faced in a real environment.
To incur some amount of time pressure to simulate a working environment,
but allow for enough time to have a reasonable attempt at seeing and briefly investigating all of the issues,
we allocate participants 30 minutes
followed by however long needed to answer a final questionnaire.
Participants were awarded online vouchers for their time regardless if they completed the experiment. Our protocol was reviewed and approved by our university's ethical committee.

\textbf{Issues}
We add 14 issues to a tracker.
7 issues each focused on security and functionality.
The security issues were created by looking at top Mitre issues%
\exthref{https://cwe.mitre.org/top25/archive/2020/2020_cwe_top25.html},
with a corresponding entry in the \texttt{find-sec-bugs} library%
\exthref{https://find-sec-bugs.github.io/bugs.htm}.
The functionality issues were created by considering what a financial app would need to offer, and potential requirements that may be faulty.
For example, one issue concerned ``Improper input validation'', with consideration to security and CWE-20.
During the experiment issues are not explicitly named as being security issues or not.
For each of these issues, the following is given: a title, a description, and a code snippet.
The issues were designed to include all the information relevant to the business,
and which would allow for full CVSS analysis.
When prioritising these issues, there will be security and functionality considerations for all,
however some issues will be more or less \emph{security critical} or \emph{business critical}.

\begin{figure}
    \centering
    \includegraphics[width=0.4\textwidth,trim={0 30mm 0 0},clip]{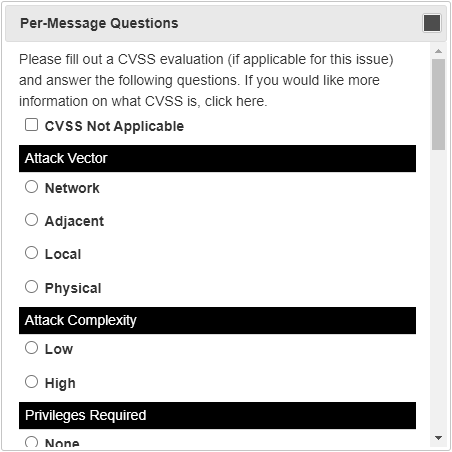}
    \caption{Screenshot of the issue questionnaire interface}%
    \label{fig:cvss}
\end{figure}

\textbf{Experiment platform}
For this experiment, we worked with a customised GitLab server.
We chose GitLab as an easily self-hostable DevOps platform we could customise for our experiment.
We added custom questionnaires to this server:
an easy-to-use drag and drop interface for choosing relative priorities,
and a questionnaire shown next to issues with CVSS and other questions (shown in Figure~\ref{fig:cvss}).
These are shown as an overlay so that the primary activity of issue analysis is always present.
The drag and drop interface
ensures that all of the prioritisation was conducted in a relative fashion
forcing a choice and preventing giving the same priority to issues.
Although we build on the GitLab interface for our study,
we are exploring the approaches developers take generally rather than specifically in a GitLab context.

\subsection{Developer Study Outcomes}

\newcommand{\sdfive}{\texttt{SD1}} 
\newcommand{\sdtwo} {\texttt{SD2}} 
\newcommand{\pmnine}{\texttt{PM1}} 
\newcommand{\pmone} {\texttt{PME}} 

Here we present the outcomes from our developer study.
We recruited 10 participants from computer science alumni, all of whom consented to take part.
4 started and completed the experiment,
and all of these participants identified as male.
We recruited a balance of project managers and software developers,
2 participants identifying expertise in each.
None of the participants felt their expertise in security was strong.
We name the less experienced participants \sdfive, \sdtwo, \pmnine,
and one project manager with more experience \pmone.

\textbf{Approaching prioritisation}
We viewed participants' approaches to choosing priorities through logged behaviour.
3 participants looked at most of the issues when making their prioritisations,
while \pmnine{} only looked at the details of 3 issues. 

\sdfive{} and \sdtwo{} changed their prioritisation as they read through the issues,
and \pmone{} preferred to make multiple prioritisations at once.
\pmnine, who did not look at many issues,
did not change the priorities from the default random order,
so we cannot draw concrete conclusions for that participant's priorities.
\sdfive{}, \sdtwo{}, and \pmnine{} on average changed 13 issue priorities from the initial random assignment,
which suggests they did engage well in this activity.

Looking at the final priorities chosen, we can see some trends.
There was agreement that ``CSRF or Referrer Missing'', ``SQL Injection'', ``Input Validation'' were the most important as they appeared in the top 5 highest priorities;
``Adding and remembering payees'' and ``Chequing'' and ``Currency converter'' were all placed in the lowest 4 priorities;
 ``Downloading PDF Summaries'' was the lowest priority issue.
This is interesting to see as it suggests, from our population sample,
that they want to rate security issues higher than non-security ones.
Participants never contacted each other yet chose similar priorities.

\textbf{Usefulness of CVSS}
\pmone, \sdtwo, and \pmnine{} attempted to generate CVSS scores for some issues.
We asked how comfortable participants felt using GitLab, choosing priorities, and completing the tasks.
Responses showed mixed responses with no clear trends across participant demographics.
Only \pmone{} felt comfortable with CVSS.
The participants showed evidence of critical security thinking, identifying where there were some security issues,
but suggested due to them not being readily exploitable, they could be de-prioritised.
When asked about the SSL issue, \pmone{} stated
``[it] is serious but is not obvious or easy to exploit so it is not as important,'' 
and on an SQL Injection issue, \sdfive{} comments
``[issues] that have the potential for data leaks, is given highest priority''.
Analyses like the ease of exploitation and impact on confidentiality are captured in CVSS,
so standardised questioning like used in CVSS could be a useful way to discuss issues amongst developers.
\pmone's comment about risk mirrors the view that CVSS is concerned with the severity over risk~\cite{SHH+_ISP-2021},
and, in that sense,  a framing of CVSS that better indicates context may be helpful.

\textbf{Engaging with security}
Participants feel there should be collective responsibility for security.
\sdfive{} suggests
``A sorted list of developer/security analysts etc{\dots} so i would know who to ping for'' 
is useful for issue tracking, to expedite seeking advice or guidance.
\pmnine{} suggests employing ``a dedicated cyber team to review and ensure best practice''. 

During the per-issue questions, we asked participants what aspect of an issue was most influential to their choice of priority.
\pmone{} and \sdtwo{} gave an explanation for every issue,
and \sdfive{} gave explanations for the 6 issues they prioritised highest.
\sdtwo{} and \pmone{}, a software developer and a project manager,
favoured `advice given' and `legal impact', respectively.
The most common aspect impacting a decision was when a fellow staffer
in the scenario gave advice explaining how an attack worked.
This is also backed up by one comment that mentioned that ``links'' to standards are an important part of prioritising an issue.
This suggests the importance of communication between a team and utilising relevant knowledge.
The next biggest impact is when there may be a legal impact to the business.
GDPR and ``privacy laws'' were specifically mentioned within this aspect.
Potentially we may see more priority given to security if laws surrounding secure programming and secure delivery of services are known.
The third biggest impact is when there is either an impact to customers, or a need to weigh between commercial and security critical interests.
\pmone{} described this tension in a comment on an issue saying
``this is a commercial issue but not a critical issue,'' 
and gave it a low priority, showing that even in the face of business, they felt security was more important.
Encouraging more dialogue in the form of advice from relevant parties during issue analysis,
taking into account tradeoffs required between commercial and security interests,
could help less experienced developers better engage with security decisions.

\section{Discussion and Conclusion}
We discuss our findings according to our 3 research questions.

\noindent\textbf{RQ1: How do software development issue tracking systems integrate security considerations?}
We see many large software projects reference security analyses in issues reported to their trackers.
Vulnerabilities like CVEs are often included in the body or comment of a report,
CVSS scores are likewise pasted into comments,
and labels are used to track the classification according to external sources like CWE classes.
Despite the apparent desire to reference such security measurements and analyses,
basic issue tracking interfaces do not offer any built-in fields to support security metadata,
with projects like Atlassian opting for a bot to add security information in Jira.

\noindent\textbf{RQ2: Does prompting for CVSS have the potential to be a useful interaction?}
In the parts of the experiment where we analyse CVSS we find mixed evidence that it could be a useful addition.
Participants were able to identify the tensions between prioritising security and functional interests.
They gave comments that mirror the approach taken by CVSS for conducting analysis.
When directly asked if they felt comfortable with CVSS, only 1 out of 4 participants felt comfortable,
so this could suggest that CVSS alone may not be suitable or would require training to increase confidence before use.
Though our sample size is low,
the fact that our more experienced project manager participant felt most comfortable with CVSS
could suggest that more experienced roles are more suited to using CVSS for prioritisation,
or that CVSS is more relevant to such roles by potentially helping them liaise with software developers who actually handle the issues.
There is evidence from prior study that giving additional advice during CVSS analysis helps to provide more accurate scoring~\cite{ABFB_PEACDASP-2018},
so any future CVSS integration should come with guidance.
Alternatives such as the Exploit Prediction Scoring System (EPSS)\exthref{https://www.first.org/epss/}~\cite{jacobs2023enhancing} are explored to better suit proactive security analysis
(EPSS was in early release when our project started).

\noindent\textbf{RQ3: How can non-security experts better engage with security during project issue management?}
Experts can find the current design of DevOps tools more useful than less experienced users.
Explanations and colour are the most useful design considerations when displaying important information for security choices.
Related studies into API design~\cite{GorAcaLIaFah_CHI-2020} find benefits in involving developers when choosing such design considerations and this may benefit issue management tools also.
Developers should be supported with external knowledge where relevant.
To best engage with security discussions,
all experts should offer their advice and context,
and if possible DevOps processes should guide those making priorities towards people best able to give this advice.
Combined with security analysis from CVSS or otherwise,
this could improve security dialogue between team members.

\bibliographystyle{plain}
\bibliography{references}

\begin{thebibliography}{10}

\bibitem{ABFB_PEACDASP-2018}
Luca Allodi, Sebastian Banescu, Henning Femmer, and Kristian Beckers.
\newblock Identifying {{Relevant Information Cues}} for {{Vulnerability
  Assessment Using CVSS}}.
\newblock In {\em Proceedings of the {{Eighth ACM Conference}} on {{Data}} and
  {{Application Security}} and {{Privacy}}}, {{CODASPY}} '18, pages 119--126,
  {New York, NY, USA}, March 2018. {Association for Computing Machinery}.

\bibitem{AO_NSPW2-2020}
Debi Ashenden and Gail Ollis.
\newblock Putting the {{Sec}} in {{DevSecOps}}: {{Using Social Practice
  Theory}} to {{Improve Secure Software Development}}.
\newblock In {\em New {{Security Paradigms Workshop}} 2020}, {{NSPW}} '20,
  pages 34--44, {New York, NY, USA}, October 2020. {Association for Computing
  Machinery}.

\bibitem{BBH+_IaST-2008}
Sarah Beecham, Nathan Baddoo, Tracy Hall, Hugh Robinson, and Helen Sharp.
\newblock Motivation in {{Software Engineering}}: {{A}} systematic literature
  review.
\newblock {\em Information and Software Technology}, 50(9):860--878, August
  2008.

\bibitem{CHP+_2ISDCS-2021}
Partha~Das Chowdhury, Joseph Hallett, Nikhil Patnaik, Mohammad Tahaei, and
  Awais Rashid.
\newblock Developers {{Are Neither Enemies Nor Users}}: {{They Are
  Collaborators}}.
\newblock In {\em 2021 {{IEEE Secure Development Conference}} ({{SecDev}})},
  pages 47--55, October 2021.

\bibitem{FIR_F--FoIRaST-2019}
FIRST.
\newblock {{CVSS}} v3.1 {{Specification Document}}.
\newblock https://www.first.org/cvss/v3.1/specification-document, 2019.

\bibitem{GorAcaLIaFah_CHI-2020}
Peter~Leo Gorski, Yasemin Acar, Luigi Lo~Iacono, and Sascha Fahl.
\newblock Listen to {{Developers}}! {{A Participatory Design Study}} on
  {{Security Warnings}} for {{Cryptographic APIs}}.
\newblock In {\em Proceedings of the 2020 {{CHI Conference}} on {{Human
  Factors}} in {{Computing Systems}}}, {{CHI}} '20, pages 1--13, {Honolulu, HI,
  USA}, April 2020. {Association for Computing Machinery}.

\bibitem{HA_C&S-2015}
Hannes Holm and Khalid~Khan Afridi.
\newblock An expert-based investigation of the {{Common Vulnerability Scoring
  System}}.
\newblock {\em Computers \& Security}, 53:18--30, September 2015.

\bibitem{jacobs2023enhancing}
Jay Jacobs, Sasha Romanosky, Octavian Suciu, Benjamin Edwards, and Armin
  Sarabi.
\newblock Enhancing vulnerability prioritization: Data-driven exploit
  predictions with community-driven insights, 2023.

\bibitem{JLEF_ITDSC-2018}
Pontus Johnson, Robert Lagerstr{\"o}m, Mathias Ekstedt, and Ulrik Franke.
\newblock Can the {{Common Vulnerability Scoring System}} be {{Trusted}}? {{A
  Bayesian Analysis}}.
\newblock {\em IEEE Transactions on Dependable and Secure Computing},
  15(6):1002--1015, November 2018.

\bibitem{Maz_P2ASCCCS-2022}
Michelle Mazurek.
\newblock We {{Are}} the {{Experts}}, and {{We Are}} the {{Problem}}: {{The
  Security Advice Fiasco}}.
\newblock In {\em Proceedings of the 2022 {{ACM SIGSAC Conference}} on
  {{Computer}} and {{Communications Security}}}, {{CCS}} '22, page~7, {New
  York, NY, USA}, November 2022. {Association for Computing Machinery}.

\bibitem{MIT_-1999}
MITRE.
\newblock Overview | {{CVE}}.
\newblock https://www.cve.org/About/Overview, 1999.

\bibitem{MIT_-2006}
MITRE.
\newblock {{CWE}} - {{Common Weakness Enumeration}}.
\newblock https://cwe.mitre.org/about/index.html, 2006.

\bibitem{MIT_-2014}
MITRE.
\newblock {{CWSS}} - {{Common Weakness Scoring System}}.
\newblock https://cwe.mitre.org/cwss/cwss\_v1.0.1.html, 2014.

\bibitem{pendletonSurveySystemsSecurity2016}
Marcus Pendleton, Richard {Garcia-Lebron}, Jin-Hee Cho, and Shouhuai Xu.
\newblock A {{Survey}} on {{Systems Security Metrics}}.
\newblock {\em ACM Computing Surveys}, 49(4):62:1--62:35, December 2016.

\bibitem{SHH+_ISP-2021}
Jonathan Spring, Eric Hatleback, Allen Householder, Art Manion, and Deana
  Shick.
\newblock Time to {{Change}} the {{CVSS}}?
\newblock {\em IEEE Security \& Privacy}, 19(2):74--78, March 2021.

\bibitem{UW_PIWCSED-2016}
Akond~Ashfaque Ur~Rahman and Laurie Williams.
\newblock Software security in {{DevOps}}: Synthesizing practitioners'
  perceptions and practices.
\newblock In {\em Proceedings of the {{International Workshop}} on {{Continuous
  Software Evolution}} and {{Delivery}}}, {{CSED}} '16, pages 70--76, {New
  York, NY, USA}, May 2016. {Association for Computing Machinery}.

\bibitem{WFW+_Q-2019}
Anna Wiedemann, Nicole Forsgren, Manuel Wiesche, Heiko Gewald, and Helmut
  Krcmar.
\newblock The {{DevOps Phenomenon}}: {{An}} executive crash course.
\newblock {\em Queue}, 17(2):Pages 40:93--Pages 40:112, April 2019.

\bibitem{XWM_P1ACCSCWSC-C1-2014}
Shundan Xiao, Jim Witschey, and Emerson {Murphy-Hill}.
\newblock Social influences on secure development tool adoption: Why security
  tools spread.
\newblock In {\em Proceedings of the 17th {{ACM}} Conference on {{Computer}}
  Supported Cooperative Work \& Social Computing - {{CSCW}} '14}, pages
  1095--1106, {Baltimore, Maryland, USA}, 2014. {ACM Press}.

\end{thebibliography}

\end{document}